\pdfoutput=1

\documentclass[twocolumn,pra, aps,superscriptaddress,showpacs]{revtex4-1}

\usepackage[colorlinks=true,citecolor=myblue,linkcolor=myblue,urlcolor=myblue]{hyperref}
\usepackage{mathptmx}
\usepackage{subfigure}
\usepackage{dcolumn}
\usepackage{amsmath,amssymb}
\usepackage{bm}
\usepackage{color}
\usepackage{latexsym}
\usepackage{color}
\usepackage[english]{babel}
\usepackage{latexsym}
\usepackage{psfrag,graphicx}
\usepackage{graphicx}
\usepackage{epsf}
\usepackage{amsmath}
\usepackage{amssymb}
\usepackage{amsfonts}
\usepackage{bm}
\usepackage{natbib}
\usepackage{epstopdf}
\usepackage{appendix}
\usepackage[export]{adjustbox}

\definecolor{mygrey}{gray}{0.35}
\definecolor{myblue}{rgb}{0.2,0.2,0.8}
\definecolor{myzard}{cmyk}{0,0,0.05,0}
\definecolor{mywhite}{rgb}{1,1,1}
\definecolor{myred}{rgb}{1,0.,0.3}

\def\be{\begin{equation}}
\def\ee{\end{equation}}
\def\ba{\begin{align}}
\def\enda{\end{align}}
\def\bi{\begin{itemize}}
\def\ei{\end{itemize}}

 \def\ee{\mathord{\rm e}}

\def\min{\mathord{\rm min}}

 \def\ee{\mathord{\rm e}}

\def\min{\mathord{\rm min}}

\renewcommand{\ee}{{\rm e}}

\def\beq{\begin{equation}}
\def\beq{\begin{equation}}
\def\eeq{\end{equation}}

 \newcommand{\ket}[1]{|#1\rangle}
 \newcommand{\bra}[1]{\langle #1|}

 \newcommand{\ketbradif}[2]{\ket{#1}\bra{#2}}
 \newcommand{\ketbra}[1]{\ketbradif {#1}{#1}}

\begin{document}

\title[Short Title]{Enhanced quantum sensing with multi-level structures of trapped ions}

\author{N. Aharon}
\affiliation{Racah Institute of Physics, The Hebrew University of Jerusalem, Jerusalem
91904, Givat Ram, Israel}
\author{M. Drewsen}
\affiliation{Department
of Physics and Astronomy, Aarhus University, DK-8000 Aarhus C, Denmark}
\author{A. Retzker}
\affiliation{Racah Institute of Physics, The Hebrew University of Jerusalem, Jerusalem
91904, Givat Ram, Israel}

\begin{abstract}
{We present a method of enhanced sensing of AC magnetic fields. The method is based on the construction of a robust qubit by the application of continuous driving fields. Specifically, magnetic noise and power fluctuations of the driving fields do not operate within the robust qubit subspace, and hence, robustness to both external and controller noise is achieved. The scheme is applicable to either a single ion or an ensemble of ions. We consider trapped-ion based implementation via the dipole transitions, which is relevant for several types of ions, such as the $^{40}{\rm{Ca}}^{+}$, $^{88}{\rm{Sr}}^{+}$, and the $^{138}{\rm{Ba}}^{+}$ ions. Taking experimental errors into account, we conclude that the coherence time of the robust qubit can be improved by up to  $\sim 4$ orders of magnitude compared to the coherence time of the bare states. We show how the robust qubit can be utilized for the task of sensing AC magnetic fields in the range $\sim 0.1 - 100$ MHz with an improvement of $\sim 2$ orders of magnitude of the sensitivity. In addition, we present a microwave based sensing scheme that is suitable for ions with a hyperfine structure, such as the $^{9}{\rm{Be}}^{+}$,$^{25}{\rm{Mg}}^{+}$,$^{43}{\rm{Ca}}^{+}$,$^{87}{\rm{Sr}}^{+}$,$^{137}{\rm{Ba}}^{+}$,$^{111}{\rm{Cd}}^{+}$,$^{171}{\rm{Yb}}^{+}$, and the $^{199}{\rm{Hg}}^{+}$ ions. This scheme enables the enhanced sensing of high frequency fields at the GHz level.}
\end{abstract}
\maketitle

\section{Introduction}

Quantum sensing \cite{review_degen,review_dima} and metrology \cite{Giovannetti2,Giovannetti1,Bollinger1} exploit physical laws governing individual quantum
systems or correlations between systems to perform detection at the limits of precision and resolution.
Improving the precision of sensing of weak electromagnetic fields is a prime goal in this field. The limit of most quantum sensing protocols scales as  $1/\sqrt{T_{2}}$ \cite{itano,helstrom,holveo}, where $T_2$ is the coherence time.
The coherence time for many experimental platforms  is limited by ambient magnetic
field fluctuations. Consequently, dynamical decoupling methods, designed to prolong the coherence time, are incorporated into the sensing schemes in order to improve the sensing precision.

Pulsed dynamical decoupling \cite{Han,CP,CPMG} is a useful tool for
prolonging the coherence time \cite{Viola,Biercuk,Du,Lange,Ryan,Naydenov,Zhi,BarGil}. Diminishing both
external and controller noise, however, requires very rapid and composite pulse sequences
\cite{Khodjasteh,Uhrig,Souza,Yang,Farfurnik} and consequently uses a considerable amount of power \cite{kurizki}.
In addition, a major drawback of incorporating state-of-the-art pulsed dynamical decoupling in sensing schemes is that usually, the frequency of
the pulses must coincide with the frequency of the sensed field; the time interval between the pulses
should be fixed to $T=\pi/\nu$, where $\nu$ is the frequency of the signal. Hence, this approach can not be used to integrate
dynamical decoupling in the sensing of high frequency fields.

Sensing of high frequency signals is of great importance, especially in the case of classical fields sensing \cite{Chipaux,Kolkowitz,loncar1,loncar2}, in the detection of electron spins in solids \cite{Sushkov,Hall2016,Dominik} and NMR \cite{Kimmich}. As the method of choice for this regime is relaxometry, the sensitivity is limited by the coherence time, and specifically by the pure dephasing time, $T_{2}^{*}$, since no dynamical decoupling schemes are employed.

Continuous dynamical decoupling \cite{Fanchini,Bermudez1,Bermudez2,Cai,Xu,Golter,rabl2009,kurizki,Jens2010,CaiCon,Itsik1,pathrick}
provides a different approach for achieving robustness to both external and controller noise \cite{Christof,ADR}.
Ultra-sensitive sensing that is based on continuous dynamical decoupling was demonstrated in \cite{Baumgart}, utilising a four  level configuration. Remarkably, continuous dynamical decoupling can be elegantly incorporated in the sensing of high frequency fields \cite{Lambda,Stark}

\begin{figure}[t]
\centering{}\includegraphics[width=0.50\textwidth]{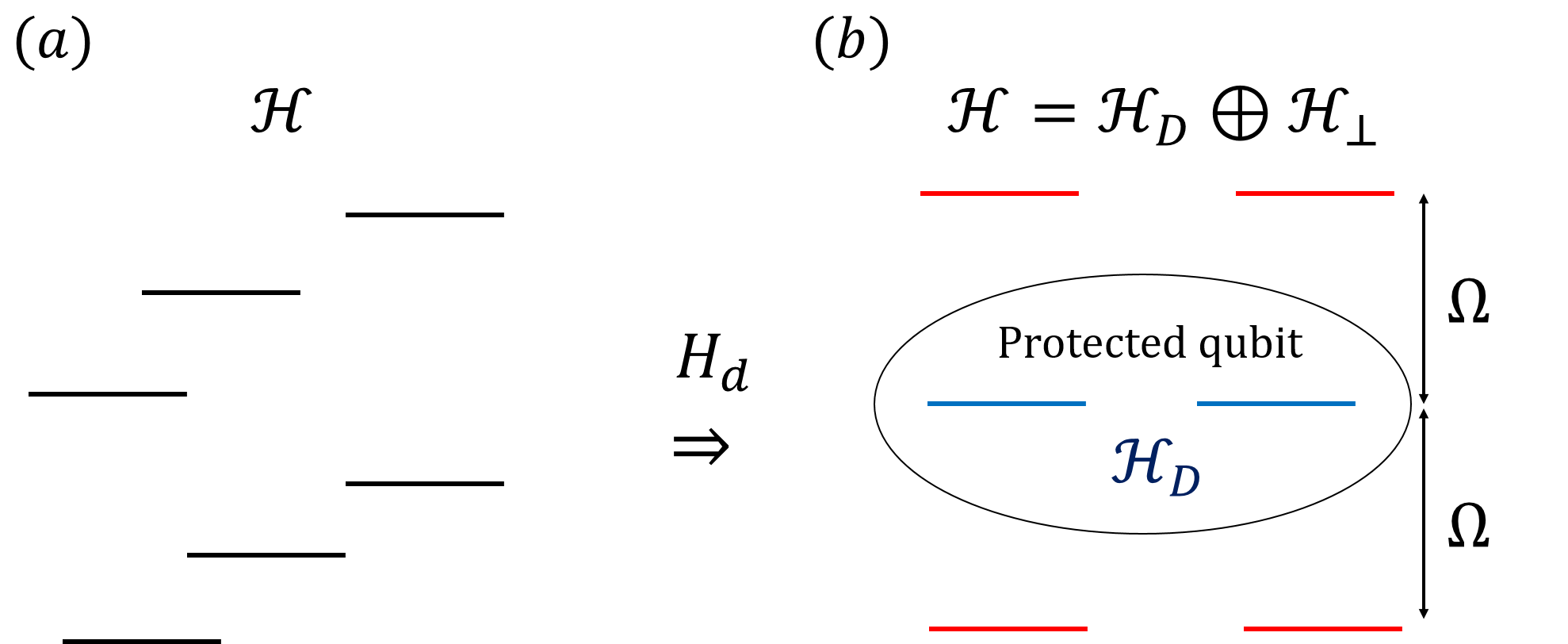}\protect\caption{\textbf{Protected qubit subspace}. By the application of continuous driving fields we create a protected qubit subspace. Magnetic noise and power fluctuations of the driving fields do not operate within the protected qubit subspace. (a) Bare states, $H_d$  (driving Hamiltonian). (b) Protected qubit subspace (blue), $\Omega$ (smallest energy gap between the robust qubit states and  non-robust states.}
\label{FR-0}
\end{figure}

\begin{figure}[t]
\includegraphics[width=0.5\textwidth,right]{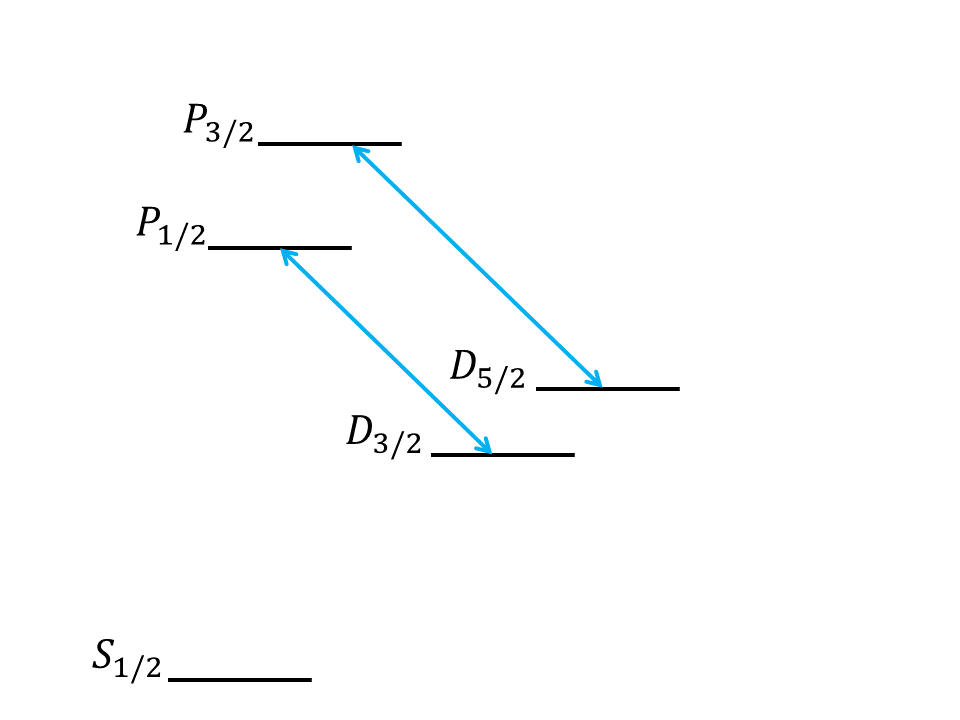}
\caption{\textbf{Typical Level structure of the , $\bf{^{40}{\rm{Ca}}^{+}}$, $\bf{^{88}{\rm{Sr}}^{+}}$, and the $\bf{^{138}{\rm{Ba}}^{+}}$ ions}. The $D_{3/2}$ subspace, which has a lifetime of $\gtrsim 1$ second, serves as the protected subspace. The  $D_{3/2}\leftrightarrow P_{1/2}$ transitions are used for the construction of the protected subspace. A similar construction can be implemented with the  $D_{5/2}$ subspace and the $D_{5/2}\leftrightarrow P_{3/2}$  transitions.}
\label{Ion}
\end{figure}

In this paper, we elaborate on the work presented in \cite{ADR} with the focus on AC magnetic field sensing. We show that utillizing a compact constuction of a protected qubit subspace that requires only two (detuned) optical driving fields (compared to five in \cite{ADR}), results in a protected qubit that can indeed be used as a sensor of AC magnetic fields in the range $\sim 0.1 - 100$ MHz with enhanced sensitivity. In addition, we present a microwave based  sensing scheme that is suitable for ions with a hyperfine structure. In this scheme a robust qubit is constructed within the ground state manifold of the ion. We show that this construction enables the enhanced sensing of high frequency fields, where the frequency of the signal corresponds to a transition between two bare states from two different hyperfine levels, which is of the order of the hyperfine splitting. The sensed frequency, at the GHz level, is tuned by the static magnetic field.

The paper is organized as follows. We begin in section \ref{PQ} with a general definition of a protected qubit subspace. In section \ref{Implementation} we discuss the realization of a protected qubit subspace with trapped ions. We start with  a review of the on-resonance construction \cite{ADR}, and then present the new compact construction. In section \ref{Serrors} we analyze the performance of the compact construction under realistic experimental errors and sources of noise. We continue with suggestions for other possible constructions in section \ref{Alternative}, and for completeness, in section \ref{Manipulation} we show how the protected qubit can be manipulated \cite{ADR}. In section \ref{Sensing} we discuss the implications of the scheme for sensing AC fields, and in section \ref{HF} we present the new microwave based construction. Finally, we end in section \ref{Conclusions} with the conclusions. \\

\section{Protected qubit subspace}\label{PQ}

We start with an explicit definition of a protected qubit subspace  \cite{ADR}. Let us  denote the protected qubit states by $\left\{ \left|D_{i}\right\rangle \right\} $ .
In what follows $H_{d}$ is the (continuous) driving Hamiltonian, $\mathcal{H}_{D}$ is the Hilbert subspace of the protected qubit,
and $\mathcal{H}_{\perp}$ is the complementary Hilbert space, that is, $\mathcal{H}=\mathcal{H}_{D}\oplus\mathcal{H}_{\perp}$.
We define the protected qubit subspace  by (See Fig. \ref{FR-0})

\begin{eqnarray}
\left\langle D_{i}\right|J_{z}\left|D_{j}\right\rangle =0 &  & \qquad\forall i,j, \label{FRD:1} \\
H_d \ket{D_i} = \lambda^{D} \ket{D_i} &  & \qquad\forall i.\label{FRD:2}
\end{eqnarray}
The first equation ensures that magnetic noise does not operate within
the protected qubit subspace; the noise can only cause transitions between
a protected state and a state in the complementary
subspace. We assume (by construction) that the energy of all states in $\mathcal{H}_{D}$ is far from the energy
of the states in $\mathcal{H}_{\perp}$. More specifically, we assume that $\nu=\min_{i} |\lambda_{i}^{\perp}-\lambda^{D}|$, where
$\lambda^{D}$ ($\lambda_{i}^{\perp}$) is an eigenvalue of an eigenstate in $\mathcal{H}_{D}$ ($\mathcal{H}_{\perp}$), is much larger
than the characteristic frequency of the noise, as in this case the
lifetime $T_{1}$ would be inversely proportional to the power spectrum of the noise  at  $\nu$. This ensures that
the rate of transitions from $\mathcal{H}_{D}$ to $\mathcal{H}_{\perp}$ due to magnetic noise is
negligible.

The second equation indicates that the protected states do not collect
a relative dynamical phase due to $H_{d}$, and are therefore  immune to noise
originating from $H_{d}$. Power fluctuations of the driving fields result in  identical
energy fluctuations of the protected states.

To summarize, the first equation ensures that the protected states
are immune to external noise, while the second equation ensures that
the protected states are also immune to controller noise.

\section{Implementation with trapped-ions}\label{Implementation}

\begin{figure}[t!]
\includegraphics[width=0.5\textwidth,left]{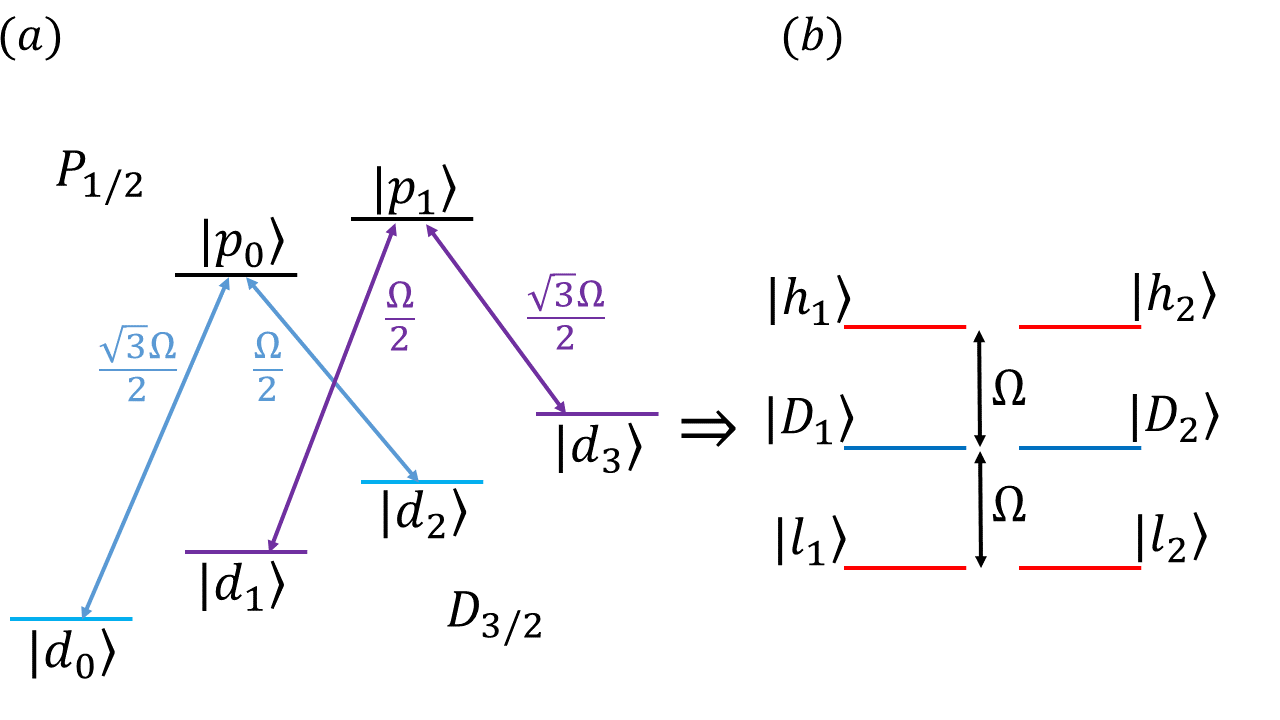} 
\protect\caption{\textbf{Construction of a protected qubit subspace}. (a) Driving fields of the two $\Lambda$ systems in the basis of the bare states. (b) Resulting eigenstates (the dressed states), where $\ket{h_{1}}=\ket{B_{1}}+\ket{p_{1}}$, $\ket{l_{1}}=\ket{B_{1}}-\ket{p_{1}}$, $\ket{h_{2}}=\ket{B_{2}}+\ket{p_{0}}$, $\ket{l_{2}}=\ket{B_{2}}-\ket{p_{0}}$, with $\ket{B_{1}}=\frac{\sqrt{3}}{2}\ket{d_{3}} +\frac{1}{2}\ket{d_{1}}$ and  $\ket{B_{2}}=\frac{\sqrt{3}}{2}\ket{d_{0}} +\frac{1}{2}\ket{d_{2}}$. The dark states $\ket{D_{1}}$ and $\ket{D_{2}}$ form the protected qubit subspace.}
\label{DS1}
\end{figure}

We consider ions such as the  $^{40}{\rm{Ca}}^{+}$, $^{88}{\rm{Sr}}^{+}$, and the $^{138}{\rm{Ba}}^{+}$ ions, that have a typical level structure, as shown in Fig. \ref{Ion}.  Because the lifetime of the $D_{3/2}$ states is $\gtrsim 1$ second, we utilize their subspace for the construction of the protected qubit subspace. For simplicity we will use the notation $\left|d_{3/2+m_{i}}\right\rangle\equiv\left|D_{3/2};m_{i}\right\rangle$, $\left|p_{1/2+m_{i}}\right\rangle\equiv\left|P_{1/2};m_{i}\right\rangle$ and $\left|s_{1/2+m_{i}}\right\rangle\equiv\left|S_{1/2};m_{i}\right\rangle$. The states,
\begin{eqnarray}
\ket{D_{1}}  & = & \frac{\sqrt{3}}{2}\ket{d_{1}} -\frac{1}{2}\ket{d_{3}} ,\\
\ket{D_{2}}  & = & \frac{\sqrt{3}}{2}\ket{d_{2}} -\frac{1}{2}\ket{d_{0}} ,
\end{eqnarray}
fulfill the first condition of a protected qubit subspace, which is given by Eq. \ref{FRD:1}, $\left\langle D_{j}\right|J_{z}\left|D_{i}\right\rangle =0$. The second condition, which is given by Eq. \ref{FRD:2}, is fulfilled by constructing the driving  fields such that $H_{d}\ket{D_{i}} =0$ ($\lambda^{D}=0$). This can be obtained by coupling the (bare) $D_{3/2}$ states to the $P_{1/2}$ states on resonance in two $\Lambda$ configurations (see Fig. \ref{DS1}).

The driving Hamiltonian of the two $\Lambda$ systems (in the interaction picture (IP) with respect to the energies of the bare states, and taking the rotating-wave approximation (RWA)) is given by
\begin{eqnarray}
H_{d} & = & \left(\frac{\Omega}{2}\ket{p_{1}} \bra{d_{1}} + \frac{\sqrt{3}\Omega}{2}\ket{p_{1}}\bra{d_{3}} \right.\nonumber \\
          & + & \;\;\; \left.\frac{\Omega}{2}\ket{p_{0}} \bra{d_{2}} + \frac{\sqrt{3}\Omega}{2}\ket{p_{0}}\bra{d_{0}} \right)+h.c.,
\label{driving}
\end{eqnarray}
where $\Omega$ is the Rabi frequency of the driving fields. The states $\left|D_{1}\right\rangle =\frac{\sqrt{3}}{2}\left|d_{1}\right\rangle -\frac{1}{2}\left|d_{3}\right\rangle $
and $\left|D_{2}\right\rangle =\frac{\sqrt{3}}{2}\left|d_{2}\right\rangle -\frac{1}{2}\left|d_{0}\right\rangle ,$
are the eigenstates of $H_{d}$ with a zero eigenvalue. The eigenvalues of the remaining four eigenstates are equal to $\pm\Omega$ (See Fig. \ref{DS1}).

\begin{figure}[t]
\centering{}
\includegraphics[width=0.50\textwidth,right]{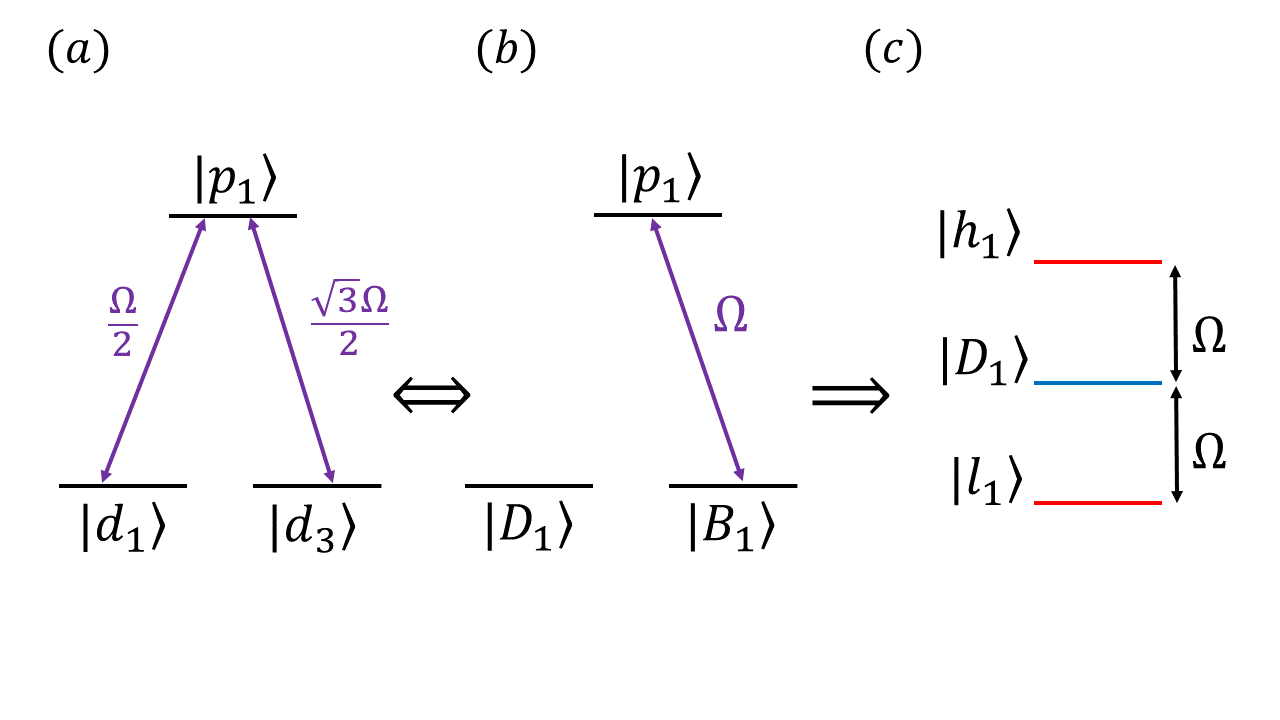}
\protect\caption{\textbf{Decoupling of a protected qubit state}. (a) Driving fields in the basis of the bare states. (b) Only the $\ket{B_{1}}$ state is coupled to the excited $\ket{p_{1}}$ state. (c) An energy gap is opened between the protected state and all other eigenstates (here in the basis of the dressed states).}
\label{DS2}
\end{figure}

The construction can be elucidated by considering one of the
$\Lambda$ systems. In the basis $\{\left|D_{1}\right\rangle =\frac{\sqrt{3}}{2}\left|d_{1}\right\rangle -\frac{1}{2}\left|d_{3}\right\rangle ,\left|B_{1}\right\rangle =\frac{1}{2}\left|d_{1}\right\rangle +\frac{\sqrt{3}}{2}\left|d_{3}\right\rangle ,\left|p_{1}\right\rangle \}$
the driving Hamiltonian of the $\Lambda$ system is given by
\begin{equation}
 H_{d}  =  \Omega\left(\left|p_{1}\right\rangle \left\langle B_{1}\right|+\left|B_{1}\right\rangle \left\langle p_{1}\right|\right).
\end{equation}
Hence, the dark state $\left|D_{1}\right\rangle $ is decoupled from the bright state \textbf{$\left|B_{1}\right\rangle $}, and the excited
state $\left|p_{1}\right\rangle $ (See Fig. \ref{DS2}). In the basis of the dressed states we have that
\begin{equation}
 H_{d} = 0\left| D_{1}\right\rangle \left\langle D_{1}\right| + \Omega\left(\left| h_{1}\right\rangle \left\langle h_{1}\right|-\left| l_{1}\right\rangle \left\langle l_{1}\right|\right),
\end{equation}
where $\ket{h_1}=\frac{1}{\sqrt{2}}\left(\ket{B_1}+\ket{p_1}\right)$, and $\ket{l_1}=\frac{1}{\sqrt{2}}\left(\ket{B_1}-\ket{p_1}\right)$.\\

\begin{figure}[t]
\centering{}
\includegraphics[width=0.50\textwidth]{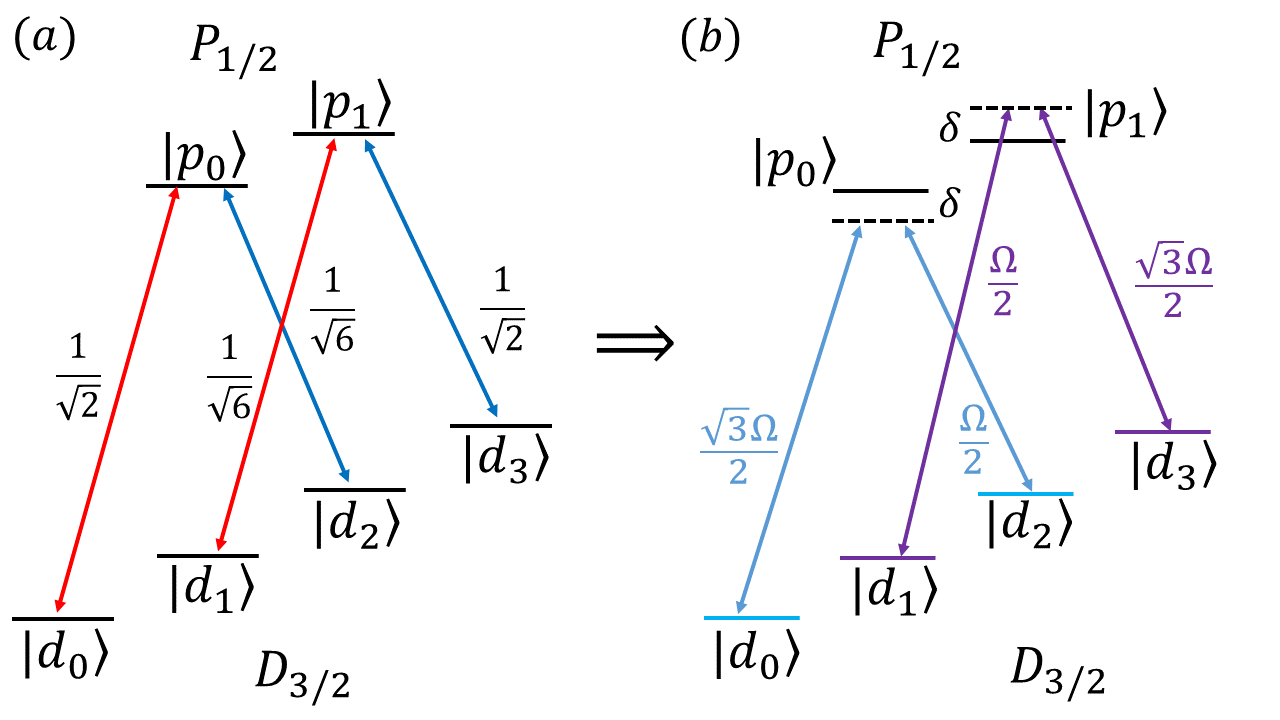}
\protect\caption{\textbf{Compact costruction}. (a) Clebsch-Gordan coefficients. (b) The ratios between the Clebsch-Gordan
coefficients of each $\Lambda$ system imply that only two driving fields with equal Rabi frequencies are required in order to construct
both protected qubit states, $\left|D_{1}\right\rangle $ and $\left|D_{2}\right\rangle$.}
\label{DS3}
\end{figure}

While for the first $\Lambda$ system the ratio between the Rabi frequencies of the $\sigma^{-}$
and $\sigma^{+}$ fields is $\frac{\Omega_{-}^{1}}{\Omega_{+}^{1}}=\sqrt{3}$, for the second $\Lambda$ system the ratio between the Rabi frequencies of the $\sigma^{-}$ and $\sigma^{+}$ fields is $\frac{\Omega_{-}^{1}}{\Omega_{+}^{1}}=\frac{1}{\sqrt{3}}$. A priori, it therefore seems that four driving fields are required, two fields for each $\Lambda$ system. However, the Clebsch-Gordan coefficients, $C_{ij}$, of the $\left|d_{i}\right\rangle \leftrightarrow\left|p_{j}\right\rangle$ transitions, fulfill the following relations (See Fig. \ref{DS3}).
\begin{equation}
\frac{C_{31}}{C_{11}}=\frac{C_{00}}{C_{20}}=\sqrt{3}.
\label{CJ}
\end{equation}
This means that only two driving fields (a $\sigma^{-}$ field and a $\sigma^{+}$ field) with an equal Rabi frequency are required for the construction of the two protected qubit states. In this case, in the first $\Lambda$ system we have a (blue) one-photon detuning and in the second system of $\left|D_{2}\right\rangle $ we have a (red) one-photon detuning of $\left|\delta\right|=\frac{1}{15}\mu_{B}B$. These one-photon detunings will not affect the protected qubit states and will only modify the eigenstates in $\mathcal{H}_{\perp}$ and their energies, which will be slightly lower. For a static magnetic field, such that $g\mu_{B}B\sim0.1-1$ MHz (which should be large enough to mitigate $\sigma_{x}$ and $\sigma_{y}$ noise), and a Rabi frequency of $\Omega\simeq 2\pi\times100$ MHz ($\Omega\gg\delta$), the protected qubit states remain well decoupled from the bright states.
In this case the dressed states $\ket{h_i}$ and $\ket{l_i}$ are modified with an amplitude mixing of $\sim \frac{\delta}{\Omega}$ between them, and their energies are shifted by an energy shift of $\sim \frac{\delta^2}{\Omega}$.  In appendix \ref{Aconst} we give a detailed description of the protected qubit construction. We consider this compact construction of the protected qubit subspace in subsequent sections.

\section{Error estimation} \label{Serrors}

In this section we consider possible experimental errors and sources of noise, and analyze their effect on the lifetime and the coherence time of the protected qubit.

\subsection{Noise and systematic shifts of the magnetic field}
Because the magnetic noise couples between a protected state and a non-protected state, in first order, the noise induces a longitudinal relaxation (decay) rate of $\sim S_{BB}(\Omega)$, where  $S_{BB}$ is the power spectrum of the noise. A large enough $\Omega$ ensures that the longitudinal relaxation rate is negligible ($ S_{BB}(\Omega) \ll \frac{1}{T_{1}}$). In our construction we consider a Rabi frequency of $\Omega \simeq 2 \pi \times 100$ MHz, which implies that for a typical magnetic noise $S_{BB}(\Omega)$ is negligible.

In second order, a Zeeman shift of $\Delta b$ results in an energy gap  of $\frac{8 }{125}\frac{ g \mu_{B}B \Delta b^2}{\Omega^2}$ between the protected qubit states (see Appendix \ref{Aerrors}). For $g\mu_B B \sim 0.01 \Omega$ the energy gap is $\sim 0.001 \frac{\Delta b^2}{\Omega}$. Even with $\Delta b \sim 50$ kHz, which corresponds to a strong magnetic noise ($T_{2}^{*}\sim 20\; \mu$s) or a large systematic shift, this implies a limit of an improvement of $\sim 6$ orders of magnitude in the coherence time.

In addition, the Zeeman shift of $\Delta b$, creates a two photon detuning, which results in an amplitude mixing between a protected state $\left|D_{i}\right\rangle $
and an excited state $\left|p_{i}\right\rangle$, where the probability to populate the excited states is $\sim (\frac{\Delta b}{\Omega})^{2}$  (See Fig. \ref{errors}). This implies a limit on the lifetime of the protected qubit due to the strong decay rate of the $P_{1/2}$ states, $\Gamma \sim 10$ MHz. For $\Delta b \sim 50$ KHz and $\Omega\simeq 2 \pi \times 100$ MHz the lifetime is limited to  $T_{1} \approx \left( \Gamma  (\frac{\Delta b}{\Omega})^{2}   \right)^{-1} \simeq 0.1$ sec, which corresponds to a limit of $4$ orders of magnitude improvement of the coherence time. Stronger driving fields of $\Omega \simeq 2 \pi \times 1$ GHz result in $T_{1} \simeq 10$ sec, which corresponds to a limit of $6$ orders of magnitude improvement in the coherence time.

\subsection{Relative amplitude error}

\begin{figure}[t]
\centering{}
\includegraphics[width=0.50\textwidth]{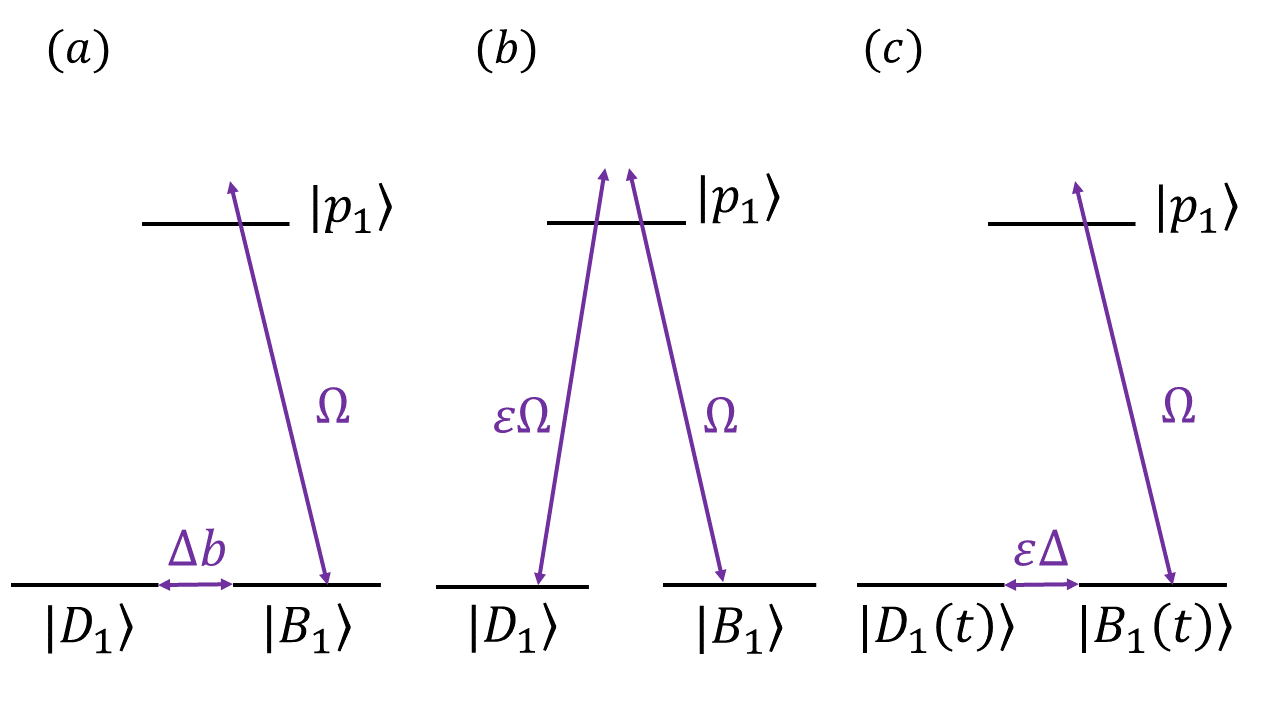}
\protect\caption{\textbf{Experimental errors}. (a) Uncertainty in the static magnetic filed couples between the $\ket{D_{i}}$ and $\ket{B_{i}}$ states, and results in a probability of $\sim (\frac{\Delta b}{\Omega})^{2}$ to populate the excited state. (b) Relative amplitude error, which results in an amplitude mixing of $\sim \varepsilon$ between the $\ket{D_{i}}$ and $\ket{B_{i}}$ states. (c)  Polarization error implies an effective coupling rate of $\varepsilon \Delta$ between the $\ket{D_{i}(t)}$ and $\ket{B_{i}(t)}$ states, and hence a probability of $\sim\left(\frac{\varepsilon\Delta}{\Omega}\right)^{2}$ to populate the excited state.}
\label{errors}
\end{figure}

In a relative amplitude error, the amplitude ratio between the two fields
(of the $\Lambda$ systems) is larger/smaller then its ideal value,
and for one of the fields we have that $\Omega\rightarrow\Omega\pm\varepsilon\Omega$.
As a result, the amplitudes of the bare states in $\left|D_{i}\right\rangle $
are modified by $\sim\varepsilon$. That is, a relative amplitude
error results in a mixing between a protected state $\left|D_{i}\right\rangle $
and the bright state $\left|B_{i}\right\rangle $ (See Fig. \ref{errors}). In
this case, the probability of being in the excited state $\left|p_{i}\right\rangle $
is zero, but because $\ket{D_{i}}$ is modified to $\left|\tilde{D}_{i}\right\rangle \approx\sqrt{1-\varepsilon^{2}}\left|D_{i}\right\rangle \pm\varepsilon\left|B_{i}\right\rangle $, $\left\langle \tilde{D}_{i}\right|\sigma_{z}\left|\tilde{D}_{i}\right\rangle $
scale as $\sim\varepsilon$. However, while the coupling rate between $\left|D_{1}\right\rangle $
and $\left|p_{1}\right\rangle $ is $\sim\varepsilon\Omega$, the coupling rate between $\left|D_{2}\right\rangle $ and
$\left|p_{2}\right\rangle $ is $\sim-\varepsilon\Omega$. As $\left\langle D_{1}\right|\sigma_{z}\left|B_{1}\right\rangle =-\left\langle D_{2}\right|\sigma_{z}\left|B_{2}\right\rangle $
, we have that the uncertainty of the energy gap between the two robust
states scales as $\left\langle \tilde{D}_{1}\right|\sigma_{z}\left|\tilde{D}_{1}\right\rangle -\left\langle \tilde{D}_{2}\right|\sigma_{z}\left|\tilde{D}_{2}\right\rangle \sim\varepsilon^{2}$ (See Appendix \ref{Aerrors}).
This limits the coherence time to $\sim\frac{T_{2}^{*}}{\varepsilon^{2}}.$
An $\varepsilon\sim10^{-2}$ implies a limit of an improvement of $4$ orders of magnitude in
the coherence time.

\subsection{Polarization errors}

Here we assume that a fraction ($\epsilon$) of a $\sigma^{-}$ beam is a $\sigma^{+}$
beam and vice versa. Hence, we have that (for the $\left|D_{1}\right\rangle $
state for example) the Rabi frequency of the $\sigma^{-}$ field is
$\Omega^{-}\left(t\right)=\left(1-\varepsilon\right)\frac{\Omega}{2}+\varepsilon e^{-i\Delta t}\frac{\Omega}{2}$,
and the Rabi frequency of the $\sigma^{+}$ field is $\Omega^{+}\left(t\right)=\left(1-\varepsilon\right)\frac{\sqrt{3}\Omega}{2}+\varepsilon e^{+i\Delta t}\frac{\sqrt{3}\Omega}{2}$
, where $\Delta=2 g_d\mu_{B}B$ and $g_d$ is the Land\'{e} g factor of the $D_{3/2}$ level. In this case, the dark state of the
$\Lambda$ system is a time dependent state, which is given (up to
normalization) by $\left|\tilde{D}_{i}\left(t\right)\right\rangle =\Omega^{-}\left(t\right)\left|d_{1}\right\rangle -\Omega^{+}\left(t\right)\left|d_{3}\right\rangle .$
Due to the time dependence of the Rabi frequencies, $\Omega^{-}\left(t\right)$
and $\Omega^{+}\left(t\right)$, there is an effective coupling rate
between the instantaneous states $\left|D_{i}\right\rangle $ and
$\left|B_{i}\right\rangle $, which is $\sim\varepsilon\Delta$. This
results in a non-zero probability to be in the excited
state $\left|p_{i}\right\rangle $, which is $\sim\left(\frac{\varepsilon\Delta}{\Omega}\right)^{2}$.
For $\varepsilon=0.5\%$ and $g\mu_{B}B\simeq0.01\Omega$, $\left(\frac{\varepsilon\Delta}{\Omega}\right)^{2}\simeq 10^{-9}$, and therefore does not limit the lifetime.
In addition, we also have that $\left\langle \tilde{D}_{i}(t)\right|\sigma_{z}\left|\tilde{D}_{i}(t)\right\rangle =0$ , and the coherence time is not affected. For more details, see Appendix \ref{Aerrors}.

\section{Alternative constructions}\label{Alternative}
There are two additional possibilities for the construction of the protected qubit subspace with optical fields.
The first alternative is to couple the $D_{3/2}$ states to the $S_{1/2}$ states instead of the $P_{1/2}$ states.
In this case the driving fields corresponds to two $V$ configurations, but the resulting protected qubit states, $\ket{D_1}$ and $\ket{D_2}$ are the same.
The advantage of this construction is that here, experimental errors can lead to a probability of populating the $S_{1/2}$ states rather than the $P_{1/2}$ states, which does not limit the lifetime of the protected qubit. However, this requires a sufficiently strong coupling rate between the $S_{1/2}$ states and the $D_{3/2}$ states such that an energy gap of $\sim 100$ kHz is formed between the protected qubit states and non-protected states.

Yet another alternative is to use the $D_{5/2}$ subspace as the protected qubit subspace. Similar to the constructions with the $D_{3/2}$ states, the Clebsch-Gordan coefficients of the $D_{5/2} \leftrightarrow P_{3/2}$ transitions imply that only two driving fields (a $\sigma^{-}$ field and a $\sigma^{+}$ field) with an equal Rabi frequency are required for the construction of the two protected qubit states. As for the $D_{3/2}$ construction, this configuration results in the two dark (zero eigenvalue) eigenstates $\ket{D_1}=\frac{1}{4}\ket{d_0}-\frac{5}{2 \sqrt{2}}\ket{d_2}+\frac{\sqrt{5}}{4}\ket{d_4}$ and   $\ket{D_2}=\frac{\sqrt{5}}{4}\ket{d_1}-\frac{5}{2 \sqrt{2}}\ket{d_3}+\frac{\sqrt{1}}{4}\ket{d_5}$, where in this case $\left|d_{5/2+m_{i}}\right\rangle\equiv\left|D_{5/2};m_{i}\right\rangle$. Hence, $\ket{D_1}$ and $\ket{D_2}$ fulfill both Eq. \ref{FRD:1} and Eq. \ref{FRD:2}, and constitute a protected qubit subspace.

\section{Initialization and single qubit gates}\label{Manipulation}

\begin{figure}[t]
\centering{}\includegraphics[width=0.50\textwidth]{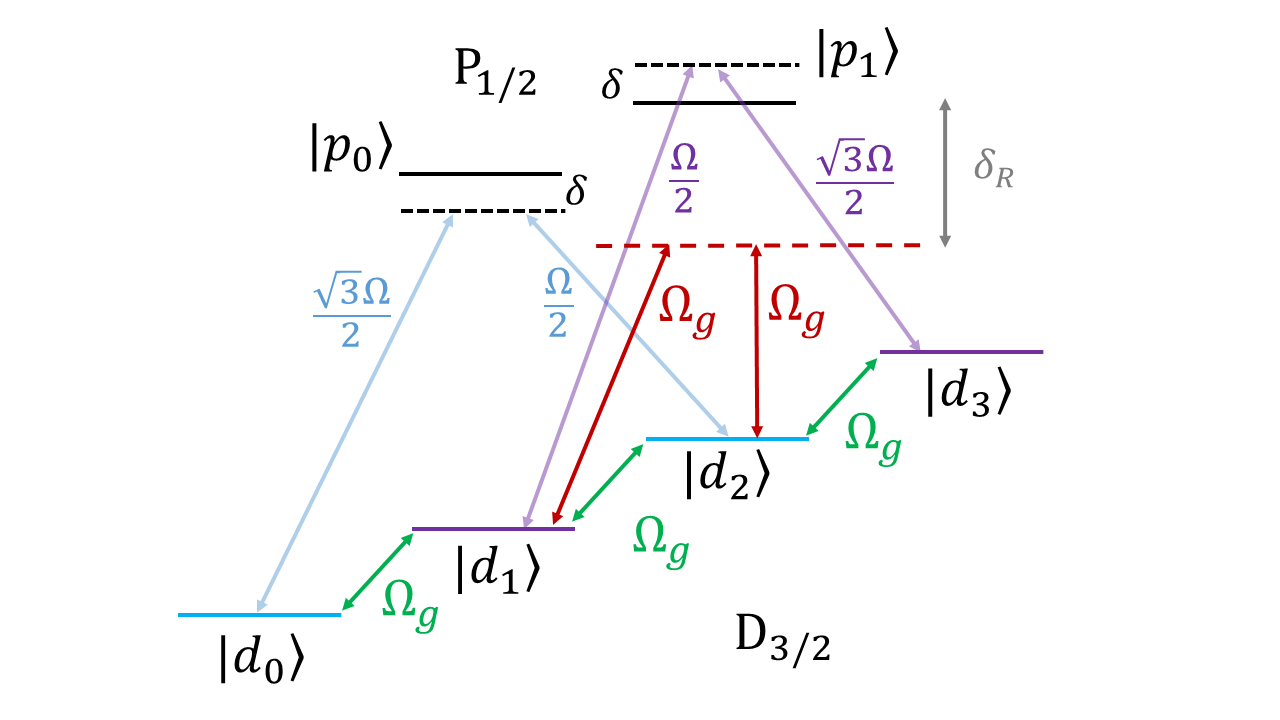}.\protect\caption{\textbf{Single qubit gates:} (i) A direct $\sigma_{y}$ gate is realized with a microwave driving on-resonance with the Zeeman splitting of the $D_{3/2}$ sub-levels (green). (ii) An effective $\sigma_{x}$ gate is obtained by a Raman transition (red).}
\label{gate}
\end{figure}

\subsection{Initialization}

Initialization may be achieved by the application of optical pumping
to the $\left|d_{3}\right\rangle $ state, followed by a STIRAP procedure,
which results in the $\left|D_{1}\right\rangle $ state. At this point,
the Rabi frequencies of the driving fields are fixed and remain so during the whole experiment. Measurements are performed
by a reversed process.

An alternative method is to perform a direct gate from an $\left|s_{i}\right\rangle $
state to one of the $\left|D_{i}\right\rangle $ states. However, this requires
two phase-matched lasers.

\subsection{Direct $\sigma_{y}$ gate}

A $\sigma_{y}$ gate can be implemented with a microwave field, which
is set to apply a $J_{y}$ gate in the bare states basis (or a $J_{x}$ with a relative phase shift of $\pi/2$), on
resonance with the (Zeeman) energy gap between the $D_{3/2}$ sub-levels
(See Fig. \ref{gate})  and corresponds to
\begin{equation}
H_g=\Omega_g \cos\left(g_d \mu_B B t\right) J_y.
\end{equation}
Moving to the IP, with respect to the bare energy gaps (and taking the RWA) leads to  $H_g^I=\frac{\Omega_g}{2} J_y$.
In the robust states basis the bare states' $J_{y}$ operates within the protected subspace $\mathcal{H}_{D}$, and within the complementary subspace $\mathcal{H}_{\perp}$, but does not couple between states in   $\mathcal{H}_{D}$ and states in $\mathcal{H}_{\perp}$.
In the protected qubit subspace $\mathcal{H}_{D}$, this results in a $\sigma_{y}$ operator which is given by
\begin{equation}
\sigma_{y}^{D}=-\frac{3i\Omega_{g}}{2}\left|D_{2}\right\rangle \left\langle D_{1}\right|+h.c..
\end{equation}

\subsection{$\sigma_{x}$ gate}

An effective $\sigma_{x}$ gate can be realized by a Raman transition
between the bare $\left|d_{1}\right\rangle $ and $\left|d_{2}\right\rangle $
states (via one of the $P_{1/2}$ states, See Fig. \ref{gate}). The (IP) Hamiltonian
of the Raman transition is
\begin{eqnarray}
H_{R} & = & \Omega_{g}\left[\left(e^{i \delta_{R} t}\ket{p_1}\bra{d_1}+h.c.\right)\right.\nonumber\\
& + & \left.\Omega_{g}\left(e^{i \delta_{R} t}\ket{p_1}\bra{d_2}+h.c.\right)\right].
\end{eqnarray}
Moving to the dressed states basis and to the IP with respect to the energies of the dressed states we obtain
\begin{eqnarray}
H_{R} & \approx & -\frac{1}{2}\sqrt{\frac{3}{2}}\Omega_{g}\left[\left(e^{i (\delta_{R}+\Omega) t}\ket{l_1}\bra{D_1}+h.c.\right)\right.\nonumber\\
& + & \left.\frac{1}{2}\sqrt{\frac{3}{2}}\Omega_{g}\left(e^{i (\delta_{R}+\Omega) t}\ket{l_1}\bra{D_2}+h.c.\right)\right]\nonumber\\
& - & \frac{1}{2}\sqrt{\frac{3}{2}}\Omega_{g}\left[\left(e^{i (\delta_{R}-\Omega) t}\ket{h_1}\bra{D_1}+h.c.\right)\right.\nonumber\\
& + & \left.\frac{1}{2}\sqrt{\frac{3}{2}}\Omega_{g}\left(e^{i (\delta_{R}-\Omega) t}\ket{h_1}\bra{D_2}+h.c.\right)\right],
\end{eqnarray}
where off-resonant coupling terms between states within $\mathcal{H}_{\perp}$ have been neglected.
Taking $\delta_{R}\gg\Omega\gg \Omega_{g}$ results in the effective Hamiltonian
\begin{equation}
\sigma_{x}^{D}\approx-\frac{3\Omega_{g}^{2}}{4\delta_{R}}\left(\left|D_{2}\right\rangle \left\langle D_{1}\right|+\left|D_{1}\right\rangle \left\langle D_{2}\right|\right),
\end{equation}
Together with the $\sigma_{y}^{D}$ gate, a
$\sigma_{z}^{D}$ gate, and hence, any single qubit unitary operation,
can be performed.\\

\begin{figure}[t!]
\centering{}\includegraphics[width=0.50\textwidth]{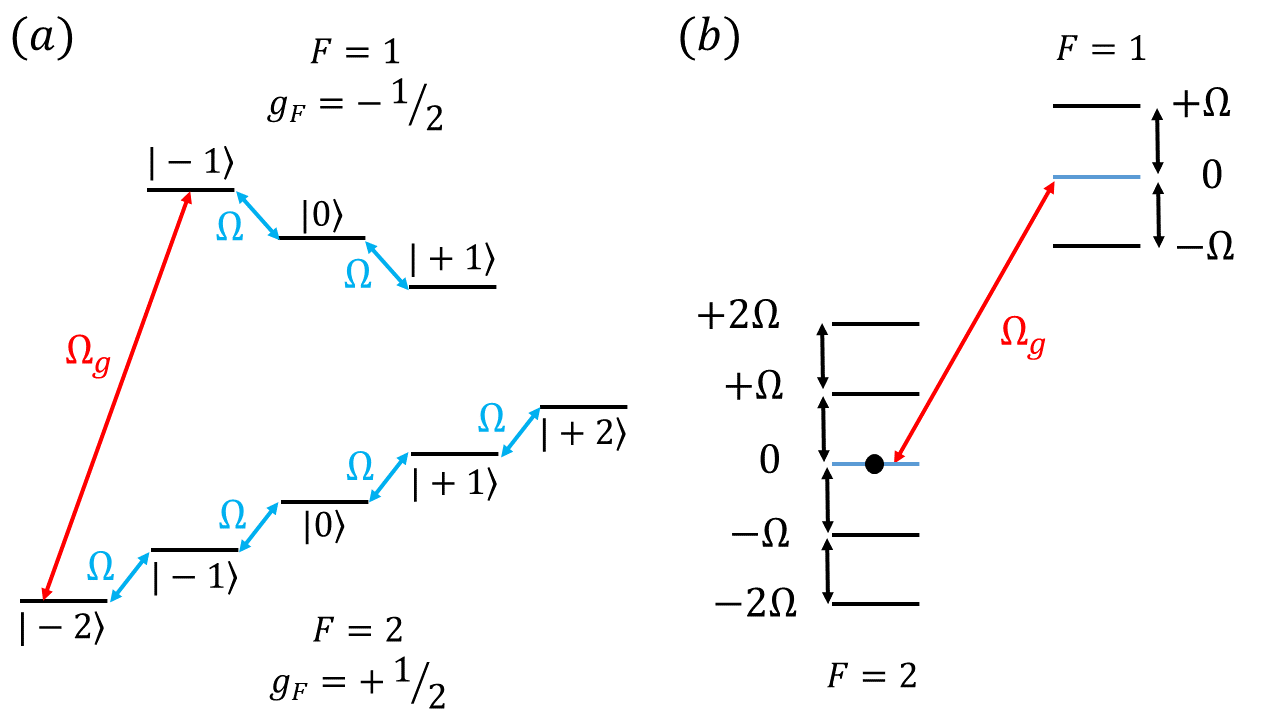}.\protect\caption{\textbf{Implementation of the scheme with an hyperfine structure:} (a) A single on resonance driving field is applied for both hyperfine levels. Here, for example, we consider $F=1$ and $F=2$ hyperfine levels of the ground state. $\Omega_{g}$ represents a signal field which is sensed by the protected qubit. Here, the frequency of the signal is on resonance with the transition between the $\ket{F=2;m_F=-2}$ and $\ket{F=1;m_F=-1}$ states. (b) Dressed states basis. We consider the $S_x$ eigenstates with the zero eigenvalue of the $F=1$ and $F=2$ levels as the protected qubit subspace. In this case the signal rotates the protected qubit state.}
\label{HF}
\end{figure}

\section{Sensing}\label{Sensing}
The operation of the $\sigma_y$ gate can be used for the task of sensing AC magnetic fields.
An AC signal whose frequency is on resonance with the Zeeman energy gap between the bare $D_{3/2}$
states will induce rotations of the protected qubit. In this case the sensitivity scales
as $\sim 1\sqrt{T_{2}}$, where $T_{2}$ is the prolonged coherence time of the protected qubit.
The range of frequencies that can be detected with this scheme is $\sim 0.1 - 100$ MHz. The
lower limit is determined by the energy gap of the bare states, which should be large enough in order to mitigate
magnetic $\sigma_x$ and $\sigma_y$ noise. The power spectrum of the noise at $\omega=g\mu_B B < 0.1$ MHz may not be negligible and hence, transverse noise may induce longitudinal relaxation of the protected qubit. The upper limit is fixed by the experimental ability to apply a strong static magnetic field. For typical ion traps, the upper bound corresponds to a Zeeman splitting of $\sim 100$ MHz. Since the phase of an external signal is usually random
with respect to the driving fields, the sensed Rabi frequency of the signal is attenuated by a factor of $1/2$.
This is due to the fact that only a $\sigma_y$ polarization induces a rotation in the protected qubit subspace.


\section{Implementation with hyperfine structure}\label{HF}
Here we consider ions with an hyperfine structure, where the ground state manifold has two hyperfine levels, $F=i$ and $F=i+1$.
For example, the $^{9}{\rm{Be}}^{+}$ and $^{137}{\rm{Ba}}^{+}$  ions have $F=1$ and $F=2$ ground states, the $^{25}{\rm{Mg}}^{+}$ ion has  $F=2$ and $F=3$, the $^{43}{\rm{Ca}}^{+}$ ion has $F=3$ and $F=4$ ground states, and the $^{87}{\rm{Sr}}^{+}$ ion has $F=4$ and $F=5$ ground states. For each ion, the Land\'{e} g-factors of the ground states are of the same magnitude, that is, $|g_{F_i}|=|g_{F_{i+1}}|$. This enables the on-resonance driving of both $F$ levels by a single driving field, and hence the construction of a protected qubit subspace  by a single on-resonance driving field. If the on-resonance driving field is polarized along the $\hat{x}$ axis, the Hamiltonian of ions with an $F=1$ and $F=2$ ground states, for example, is given by
\begin{equation}
H=g_{F_2}\mu_{B}B(F_z^2-F_z^1) + 2\Omega(F_x^2-F_x^1)\cos\left(g_{F_2}\mu_{B}B t\right),
\end{equation}
where $F_z^1$ ($F_z^2$) and $F_x^1$ ($F_x^2$) are the $F_z$ and $F_x$ operators, which operate on the $F=1$ ($F=2$) levels respectively.
In the IP we have that
\begin{equation}
H_I=\Omega(F_x^2-F_x^1),
\end{equation}
and so the dressed eigenstates of the driving field are the  $F_{x}$ eigenstates, which fulfill Eq. \ref{FRD:1} and are robust to magnetic noise. The eigevalues of the dressed $F=1$ and $F=2$ states are $\{-\Omega,0,+\Omega\}$ and $\{-2\Omega,-\Omega,0,+\Omega\,+2\Omega\}$ respectively. Choosing two $F_{x}$ eigenstates, one from each $F$ level, that have the same eigenvalue (i.e., the same energy) ensures that these two states also fulfill Eq. \ref{FRD:2}, and are therefore also robust to power fluctuations of the driving field (See Fig. \ref{HF}).

Because the protected qubit states have the same energy, their on-resonance coupling frequency is a frequency of an allowed transition between two bare states that have a non-zero amplitude component in the dressed states, and up to the Zeeman splitting corresponds to the hyperfine splitting. Denote by $\omega_{HF}$ the frequency which corresponds to the hyperfine splitting, and consider the two eigenstates with the zero eigenvalue as the protected qubit states, $\ket{D_1}=\frac{1}{\sqrt{2}}\left(\ket{+1}-\ket{-1}\right)$ and $\ket{D_2}=\frac{1}{2}\left(\sqrt{\frac{3}{2}}\ket{+2}-\ket{0}+\sqrt{\frac{3}{2}}\ket{-2}\right)$. The driving field
\begin{equation}
H_g=\Omega_g S_x\cos\left[\left(\omega_{HF}+3 g_{F_2}\mu_{B}B\right) t\right],
\end{equation}
which is on resonance with the $\ket{F=2;m_F=-2}\leftrightarrow\ket{F=1;m_F=-1}$ transition corresponds to the on resonance coupling between the two protected states. In the first IP with respect to the Zeeman splitting
\begin{equation}
H_g\approx\frac{\sqrt{3}\Omega_g}{8}\left(\ket{D_2}\bra{D_1}e^{-i \omega_{HF}t}+\ket{D_1}\bra{D_2}e^{+i \omega_{HF}t}\right),
\end{equation}
where all coupling terms within $\mathcal{H}_{\perp}$, and off-resonance coupling terms between a state in $\mathcal{H}_{D}$ and a state in $\mathcal{H}_{\perp}$ have been neglected. Moving now to the IP with respect to the hyperfine splitting results in
\begin{equation}
H_g\approx\frac{\sqrt{3}\Omega_g}{8}\left(\ket{D_2}\bra{D_1}+\ket{D_1}\bra{D_2}\right).
\end{equation}

For ions with an $F=0$ level, such as the $^{111}{\rm{Cd}}^{+}$,$^{171}{\rm{Yb}}^{+}$ and the $^{199}{\rm{Hg}}^{+}$ ions, the driving field will operate only on the $F=1$ level. The protected qubit states are then seen to be the bare $\ket{F=0;m_F=0}$ state and the dressed $F_x$ eigenstate with the zero eigenvalue of the $F=1$ level (which is the $\ket{D}=\frac{1}{\sqrt{2}}(\ket{+1}-\ket{-1})$ state.

The hyperfine ground states energy gap in these ions is $1-40$ GHz, and hence, our scheme provides a method for the enhanced sensing of high frequency signals. While with common high frequency sensing methods, the obtained sensitivity is limited by the pure dephasing time $T_{2}^{*}$, in our method the sensitivity is limited by the prolonged coherence time.

\section{Conclusions}\label{Conclusions}
This paper describes a scheme for the construction of a protected qubit subspace. Specifically, we present a simplified and compact optical implementation of the scheme, which is of relevance to several types of ions and can be used for enhanced sensing of AC magnetic fields. The scheme is applicable to either a single ion or an ensemble of ions. Taking experimental errors into account, we predict an improvement of ~2 orders of magnitude in sensitivity.
We also describe a new implementation of the scheme that is suitable for ions with hyperfine structure, where a protected qubit subspace is constructed within the ground state manifold by the application of a single microwave field. Importantly, this implementation enables the enhanced sensing of high frequency fields at the GHz level, which corresponds to the energy gap between the hyperfine levels.
Owing to the high resolution of the scheme, the implication of this method for wireless communication \cite{Baylis} may prove to be very significant. Moreover, this method could also be used to improve the coupling to quantum systems \cite{Cai}, and to improve the resolution of classical fields sensing \cite{Chipaux,Kolkowitz,loncar1,loncar2}.

\section{Acknowledgements}
A. R. acknowledges the support of the Israel Science Foundation(grant no. 1500/13), the support of the European commission (STReP EQUAM Grant Agreement
No. 323714), EU Project DIADEMS, the Marie Curie Career Integration Grant (CIG) IonQuanSense(321798), the Niedersachsen-Israeli Research Cooperation Program and DIP program (FO 703/2-1). This work is partially supported by the US Army Research Office under Contract W911NF-15-1-0250.
This project has received funding from the European Union as Horizon 2020 research and innovation programme under grant agreement No 667192.

M.D. appreciates generous support through the The Danish Council for Independent Research, Sapere Aude DFF-Advanced Grant; and the VELUX Foundations.

\appendix

\section{}\label{Aconst}
In this section we describe the construction of the protected qubit subspace in detail.

\subsection{Ideal construction}
Here we assume that for each $\Lambda$ system we have two independent driving fields that operate only on the transitions within the $\Lambda$ system.
Denote by $\omega$ and $B$ the energy gap between the $D_{3/2}$ and $P_{1/2}$ levels, and the amplitude of the static magnetic field ($B$ here corresponds to $\mu_{B} B$). Note that the Land\'{e} g-factor of the $D_{3/2}$ and $P_{1/2}$ levels is $g_{d}=\frac{4}{5}$ and  $g_{p}=\frac{2}{3}$ respectively. We then have that

\begin{eqnarray}
H_{d1}^{0} & = & (\omega+\frac{B}{3})\ketbra{p_{1}} - \frac{2B}{5}\ketbra{d_{1}} +  \frac{6B}{5}\ketbra{d_{3}},\nonumber \\
H_{d2}^{0} & = & (\omega-\frac{B}{3})\ketbra{p_{2}} - \frac{6B}{5}\ketbra{d_{0}} +  \frac{2B}{5}\ketbra{d_{2}}.
\end{eqnarray}
The Hamiltonians of the on-resonance driving fields are therefore given by

\begin{eqnarray}
H_{d1} & = & \Omega\cos\left[\left(\omega+\frac{11B}{15}\right)t\right]\left(\ket{p_{1}}\bra{d_{1}}+h.c.\right) \nonumber\\
& + & \sqrt{3}\Omega\cos\left[\left(\omega-\frac{13B}{15}\right)t\right]\left(\ket{p_{1}}\bra{d_{3}}+h.c.\right),\nonumber \\
H_{d2} & = & \Omega\cos\left[\left(\omega+\frac{13B}{15}\right)t\right]\left(\ket{p_{0}}\bra{d_{0}}+h.c.\right) \nonumber\\
& + & \sqrt{3}\Omega\cos\left[\left(\omega-\frac{11B}{15}\right)t\right]\left(\ket{p_{0}}\bra{d_{2}}+h.c.\right),
\end{eqnarray}
and the total Hamiltonian is
\begin{equation}
H_d = H_{d1}^{0} + H_{d2}^{0} + H_{d1} + H_{d2}.
\end{equation}
Moving to the IP with respect to $H^{0}= H_{d1}^{0}+ H_{d2}^{0}$,
and making the RWA we arrive at Eq. (\ref{driving}),

\begin{eqnarray}
H_{d}^{I} & = & \left(\frac{\Omega}{2}\ket{p_{1}} \bra{d_{1}} + \frac{\sqrt{3}\Omega}{2}\ket{p_{1}}\bra{d_{3}} \right.\nonumber \\
          & + & \;\;\; \left.\frac{\Omega}{2}\ket{p_{0}} \bra{d_{2}} + \frac{\sqrt{3}\Omega}{2}\ket{p_{0}}\bra{d_{0}} \right)+h.c..
\end{eqnarray}
In the basis $\{\left|D_{1}\right\rangle =\frac{\sqrt{3}}{2}\left|d_{1}\right\rangle -\frac{1}{2}\left|d_{3}\right\rangle ,\left|B_{1}\right\rangle =\frac{1}{2}\left|d_{1}\right\rangle +\frac{\sqrt{3}}{2}\left|d_{3}\right\rangle ,\left|D_{2}\right\rangle =\frac{\sqrt{3}}{2}\left|d_{2}\right\rangle -\frac{1}{2}\left|d_{0}\right\rangle ,\left|B_{2}\right\rangle =\frac{1}{2}\left|d_{2}\right\rangle +\frac{\sqrt{3}}{2}\left|d_{0}\right\rangle ,\left|p_{1}\right\rangle \,,\left|p_{0}\right\rangle \}$ the Hamiltonian is given by

\begin{equation}
H_{d}^{I}=\Omega\left(\left|p_{1}\right\rangle \left\langle B_{1}\right|+\left|p_{0}\right\rangle \left\langle B_{2}\right|\right) + h.c.,
\end{equation}
where only the bright states are coupled to the excited states. In the basis of its eigenstates, $H_{d}^{I}$ reads as
\begin{eqnarray}
 H_{d}^{I} & = & 0\left(\left| D_{1}\right\rangle \left\langle D_{1}\right|+\left| D_{2}\right\rangle \left\langle D_{2}\right|\right)\nonumber\\
& + & \Omega\left(\left| h_{1}\right\rangle \left\langle h_{1}\right|+\left| h_{2}\right\rangle \left\langle h_{2}\right|\right)\nonumber\\
& - &  \Omega\left(\left| l_{1}\right\rangle \left\langle l_{1}\right|+\left| l_{2}\right\rangle \left\langle l_{2}\right|\right),
\end{eqnarray}
 where $\ket{h_1}=\frac{1}{\sqrt{2}}\left(\ket{B_1}+\ket{p_1}\right)$, $\ket{l_1}=\frac{1}{\sqrt{2}}\left(\ket{B_1}-\ket{p_1}\right)$, $\ket{h_2}=\frac{1}{\sqrt{2}}\left(\ket{B_2}+\ket{p_0}\right)$, and $\ket{l_2}=\frac{1}{\sqrt{2}}\left(\ket{B_2}-\ket{p_0}\right)$.

\subsection{Compact construction}
The relation between the Clebsch-Gordan coefficients of the $D_{3/2} \leftrightarrow P_{1/2}$ transitions, given by Eq. \ref{CJ}, suggest that the protected qubit subspace can be constructed by only two driving fields, which operate on both $\Lambda$ systems and have the same Rabi frequency. We set the driving frequencies of the $\sigma^{+}$ and $\sigma^{-}$ fields to $\omega_{+}=\omega + g_{d} B$ and $\omega_{-}=\omega - g_{d} B$, and define  $\delta=\frac{1}{15}B$. In this case, the Hamiltonians of the driving fields are given by

\begin{eqnarray}
H_{d1} & = & \Omega\cos\left[\omega_{+} t\right]\left(\ket{p_{1}}\bra{d_{1}}+h.c.\right) \nonumber\\
& + & \sqrt{3}\Omega\cos\left[\omega_{-} t\right]\left(\ket{p_{1}}\bra{d_{3}}+h.c.\right),\nonumber \\
H_{d2} & = & \Omega\cos\left[\omega_{+} t\right]\left(\ket{p_{0}}\bra{d_{0}}+h.c.\right) \nonumber\\
& + & \sqrt{3}\Omega\cos\left[\omega_{-} t\right]\left(\ket{p_{0}}\bra{d_{2}}+h.c.\right).
\end{eqnarray}
Moving to the IP with respect to $H^{0}= H_{d1}^{0}+ H_{d2}^{0}+ \delta\left(\ketbra{p_{1}}-\ketbra{p_{0}}\right)$,
and making the RWA we obtain
\begin{eqnarray}
H_{d}^{I} & = & \delta\left(\ketbra{p_{0}}-\ketbra{p_{1}}\right)\nonumber\\
          & + & \left(\frac{\Omega}{2}\ket{p_{1}} \bra{d_{1}} + \frac{\sqrt{3}\Omega}{2}\ket{p_{1}}\bra{d_{3}} \right.\nonumber \\
          & + & \;\;\; \left.\frac{\Omega}{2}\ket{p_{0}} \bra{d_{2}} + \frac{\sqrt{3}\Omega}{2}\ket{p_{0}}\bra{d_{0}} \right)+h.c.,
\end{eqnarray}
where $\pm\delta$ is the one-photon detunings of the driving fields. We consider the regime where $B\ll\Omega$, which means that the ideal $\ket{h_i}$ and $\ket{l_i}$ eigenstates are slightly mixed, with an amplitude mixing of $\sim \frac{\delta}{\Omega}$, and their energies are shifted by an energy shift of $\sim \frac{\delta^2}{\Omega}$. The protected qubit states remain exactly the same and well decoupled from the bright and excited states.

\section{}\label{Aerrors}
In this section we derive the results presented in Sec. \ref{Serrors}.
We consider the compact construction of the protected qubit subspace.

\subsection{Magnetic shift}
A magnetic shift affects both $D_{3/2}$ and $P_{1/2}$ levels, and is described by the Hamiltonian
\begin{equation}
H_b = g_d \Delta b J_z^D + g_p \Delta b J_z^P,
\end{equation}
where, $J_z^D$ ($J_z^P$) operates on the $D_{3/2}$ ($P_{1/2}$) states.
Adding $H_b$ to $H_d$ and moving to the IP results in
\begin{eqnarray}
H_{d}^{I} & = & \left(g_p\Delta b - \delta\right)J_z^P + g_d \Delta b J_z^D \\
          & + & \left(\frac{\Omega}{2}\ket{p_{1}} \bra{d_{1}} + \frac{\sqrt{3}\Omega}{2}\ket{p_{1}}\bra{d_{3}} \right.\nonumber \\
          & + & \;\;\; \left.\frac{\Omega}{2}\ket{p_{0}} \bra{d_{2}} + \frac{\sqrt{3}\Omega}{2}\ket{p_{0}}\bra{d_{0}} \right)+h.c..
\end{eqnarray}
In second order of $\Delta b$, the energies of the protected $\ket{D_1}$ and $\ket{D_2}$ states are modified to $+\frac{4}{125}\frac{B \Delta b^2}{\Omega^2}$ and  $-\frac{4}{125}\frac{B \Delta b^2}{\Omega^2}$ respectively.

Because the magnetic shift couples between the protected (dark) states and the bright states, and the driving fields couple between the bright states and the excited states, the magnetic shift also results in a small amplitude mixing of $\sim \left(\frac{\Delta b}{\Omega}\right)^2$ between a protected state and an excited state.

\subsection{Relative amplitude}
We consider the case where  the amplitude of one of the driving fields, the $\sigma^+$ field for example, is $(1+\varepsilon)\Omega$ instead of $\Omega$. The Hamiltonian in the IP is therefore modified to
\begin{eqnarray}
H_{d}^{I} & = & \delta\left(\ketbra{p_{0}}-\ketbra{p_{1}}\right)\nonumber\\
          & + & \left(\frac{\left(\Omega+\varepsilon\right)}{2}\ket{p_{1}} \bra{d_{1}} + \frac{\sqrt{3}\Omega}{2}\ket{p_{1}}\bra{d_{3}} \right.\nonumber \\
          & + & \;\;\; \left.\frac{\Omega}{2}\ket{p_{0}} \bra{d_{2}} + \frac{\sqrt{3}\left(\Omega+\varepsilon\right)}{2}\ket{p_{0}}\bra{d_{0}} \right)+h.c.,
\end{eqnarray}
which in the $\{\ket{D_i},\ket{B_i}\}$ basis is given by
\begin{eqnarray}
H_{d}^{I} & = & \delta\left(\ketbra{p_{0}}-\ketbra{p_{1}}\right)\nonumber\\
          & + & \Omega \left(1+\frac{\varepsilon}{4}\right)\left(\left|p_{1}\right\rangle \left\langle B_{1}\right|+ h.c.\right) \nonumber \\
          & + & \Omega \left(1+\frac{3\varepsilon}{4}\right)\left(\left|p_{0}\right\rangle \left\langle B_{2}\right| + h.c.\right)\nonumber \\
          & + & \frac{\sqrt{3}\Omega\varepsilon}{4}\left(\left|p_{1}\right\rangle \left\langle D_{1}\right|+ h.c.\right)\nonumber \\
          & - & \frac{\sqrt{3}\Omega\varepsilon}{4}\left(\left|p_{0}\right\rangle \left\langle D_{2}\right|+ h.c.\right),
\end{eqnarray}
The protected states are therefore modified to
\begin{eqnarray}
\ket{\tilde{D}_{1}} & \approx & \ket{D_{1}} +  \frac{\sqrt{3}\varepsilon}{4}\ket{B_{1}}\nonumber\\
\ket{\tilde{D}_{2}} & \approx & \ket{D_{2}} -  \frac{\sqrt{3}\varepsilon}{4}\ket{B_{2}}.
\end{eqnarray}
In first order of $\varepsilon$, for both $\ket{\tilde{D}_{i}}$ states  $\left\langle \tilde{D}_{i}\right|\sigma_{z}\left|\tilde{D}_{i}\right\rangle = \frac{3\varepsilon}{4}$. In second order of $\varepsilon$ we find that $\left\langle \tilde{D}_{1}\right|\sigma_{z}\left|\tilde{D}_{1}\right\rangle -\left\langle \tilde{D}_{2}\right|\sigma_{z}\left|\tilde{D}_{2}\right\rangle =\frac{3\varepsilon^2}{4}$, which
limits the coherence time to $\sim\frac{T_{2}^{*}}{\varepsilon^{2}}.$

\subsection{Polarization error}
Here we assume that a fraction of a $\sigma^{-}$ beam is a $\sigma^{+}$
beam and vice versa. Hence, we find that, for the $\left|D_{1}\right\rangle $
state for example, the Rabi frequency of the $\sigma^{-}$ field is
$\Omega^{-}\left(t\right)=\left(1-\varepsilon\right)\frac{\Omega}{2}+\varepsilon e^{-i\Delta t}\frac{\Omega}{2}$,
and the Rabi frequency of the $\sigma^{+}$ field is $\Omega^{+}\left(t\right)=\left(1-\varepsilon\right)\frac{\sqrt{3}\Omega}{2}+\varepsilon e^{+i\Delta t}\frac{\sqrt{3}\Omega}{2}$ , where $\Delta= 2g_dB$.
The Hamiltonian in the IP and in the $\{\ket{D_i},\ket{B_i}\}$ basis is given by
\begin{eqnarray}
H_{d}^{I} & = & \delta\left(\ketbra{p_{0}}-\ketbra{p_{1}}\right)\nonumber\\
          & + & \Omega \left(1-\varepsilon+\varepsilon\cos\left[\Delta t\right]+\frac{i}{2}\varepsilon\sin\left[\Delta t\right]\right)\left(\left|p_{1}\right\rangle \left\langle B_{1}\right|+ h.c.\right) \nonumber \\
          & + & \Omega \left(1-\varepsilon+\varepsilon\cos\left[\Delta t\right]-\frac{i}{2}\varepsilon\sin\left[\Delta t\right]\right)\left(\left|p_{0}\right\rangle \left\langle B_{2}\right| + h.c.\right)\nonumber \\
          & + & \frac{i\sqrt{3}}{2}\varepsilon \Omega\sin\left[\Delta t\right]\left(\left|p_{1}\right\rangle \left\langle D_{1}\right|+ h.c.\right)\nonumber \\
          & - & \frac{i\sqrt{3}}{2}\varepsilon \Omega\sin\left[\Delta t\right]\left(\left|p_{0}\right\rangle \left\langle D_{2}\right|+ h.c.\right).
\end{eqnarray}
The modified time-dependent robust states, $\ket{\tilde{D}_{i}(t)}$ are robust to magnetic noise and satisfy Eq. \ref{FRD:1}, $\bra{\tilde{D}_{i}(t)} J_z \ket{\tilde{D}_{i}(t)}=0$. The dominant effect of polarization error is a non-zero probability to populate the excited states. Similar to adiabatic transfer, the time dependency of the driving fields results is a coupling between the instantaneous $\ket{D_{i}}$ and  $\ket{B_{i}}$ states, which is $\sim \Delta \varepsilon$. Hence the probability of populating the excited state is $\sim \left(\frac{\Delta\varepsilon}{\Omega}\right)^2$.


\begin{references}

\bibitem{review_degen} C. L. Degen, F. Reinhard, and P. Cappellaro. arXiv preprint,
arXiv:1611.02427, 2016.

\bibitem{review_dima} Dmitry Budker and Michael Romalis. \href{http://www.nature.com/nphys/journal/v3/n4/abs/nphys566.html}{Nature Physics {\bf 3,} 227 (2007)}


\bibitem{Bollinger1} JJ Bollinger, Wayne M Itano, DJ Wineland, and DJ Heinzen. \href{https://journals.aps.org/pra/abstract/10.1103/PhysRevA.54.R4649}{Phys. Rev. A {\bf 54,} R4649(R)  (1996)}


\bibitem{Giovannetti1} Vittorio Giovannetti, Seth Lloyd, and Lorenzo Maccone. \href{http://science.sciencemag.org/content/306/5700/1330}{Science, {\bf 306} 1330, (2004).}


\bibitem{Giovannetti2} Vittorio Giovannetti, Seth Lloyd, and Lorenzo Maccone. \href{https://journals.aps.org/prl/abstract/10.1103/PhysRevLett.96.010401}{Phys. Rev. Lett. {\bf 96,} 010401 (2006)}

\bibitem{itano} WM Itano, JC Bergquist, JJ Bollinger, JM Gilligan, DJ Heinzen, FL Moore, MG Raizen,
and DJ Wineland. \href{https://journals.aps.org/pra/abstract/10.1103/PhysRevA.47.3554}{Phys. Rev. A {\bf 47,} 3554}

\bibitem{helstrom} Carl W Helstrom. Quantum detection and estimation theory. Journal of Statistical Physics,
1(2):231, 1969.

\bibitem{holveo} Alexander S Holevo. Probabilistic and statistical aspects of quantum theory, volume 1. Springer
Science and Business Media, 2011.

\bibitem{Han} E. L. Hahn, \href{http://dx.doi.org/10.1103/PhysRev.80.580}{Phys. Rev. \textbf{80}, 580 (1950).}

\bibitem{CP} H. Y. Carr and E. M. Purcell, \href{http://dx.doi.org/10.1103/PhysRev.94.630}{Phys. Rev. \textbf{94}, 630 (1954).}

\bibitem{CPMG} S. Meiboom and D. Gill, \href{http://dx.doi.org/10.1063/1.1716296}{Rev. Sci. Instrum. \textbf{29}, 688 (1958).}

\bibitem{Viola} L. Viola and S. Lloyd, \href{http://dx.doi.org/10.1103/PhysRevA.58.2733}{Phys. Rev. A \textbf{58}, 2733 (1998).}

\bibitem{Biercuk} M. J. Biercuk, H. Uys, A. P. VanDevender, N. Shiga, W. M. Itano, and J. J. Bollinger, \href{http://dx.doi.org/10.1038/nature07951}{Nature \textbf{458}, 996 (2009).}

\bibitem{Du} J. Du, X. Rong, N. Zhao, Y. Wang, J. Yang and R. B. Liu \href{http://dx.doi.org/10.1038/nature08470}{Nature \textbf{461} 1265 (2009).}

\bibitem{Lange} G. de Lange, Z. H. Wang, D. Rist\`{e}, V. V. Dobrovitski, and R. Hanson \href{http://dx.doi.org/10.1126/science.1192739}{Science \textbf{330}, 60 (2010).}

\bibitem{Ryan} C. A. Ryan, J. S. Hodges, and D. G. Cory, \href{http://dx.doi.org/10.1103/PhysRevLett.105.200402}{Phys. Rev. Lett. \textbf{105}, 200402 (2010).}

\bibitem{Naydenov} B. Naydenov, F. Dolde, L. T. Hall, C. Shin, H. Fedder, L. C. L. Hollenberg, F. Jelezko, and J. Wrachtrup, \href{http://dx.doi.org/10.1103/PhysRevB.83.081201}{Phys. Rev. B \textbf{83}, 081201(R) (2011).}

\bibitem{Zhi} Z.-H. Wang, G. de Lange, D. Rist\`{e}, R. Hanson, and V. V. Dobrovitski, \href{http://dx.doi.org/10.1103/PhysRevB.85.155204}{Phys. Rev. B \textbf{85}, 155204 (2012).}

\bibitem{BarGil} N. Bar-Gill,	L.M. Pham,	A. Jarmola,	D. Budker, and R.L. Walsworth, \href{http://dx.doi.org/10.1038/ncomms2771}{Nat. Commun. \textbf{4}, 1743 (2012).}

\bibitem{Khodjasteh} K. Khodjasteh and D. A. Lidar, \href{http://dx.doi.org/10.1103/PhysRevLett.95.180501}{Phys. Rev. Lett. \textbf{95}, 180501(2005).}

\bibitem{Uhrig} G. Uhrig, \href{http://dx.doi.org/10.1103/PhysRevLett.98.100504}{Phys. Rev. Lett. \textbf{98}, 100504 (2007).}

\bibitem{Souza} A. M. Souza, G. A. \'{A}lvarez, and D. Suter, \href{http://dx.doi.org/10.1103/PhysRevLett.106.240501}{Phys. Rev. Lett. \textbf{106}, 240501 (2011).}

\bibitem{Yang} Wen Yang, Zhen-Yu Wang, Ren-Bao Liu, \href{http://dx.doi.org/10.1007/s11467-010-0113-8}{Front. Phys. \textbf{6}, 2 (2011).}

\bibitem{Farfurnik} D. Farfurnik, A. Jarmola, L. M. Pham, Z. H. Wang, V. V. Dobrovitski, R. L. Walsworth, D. Budker, and N. Bar-Gill, \href{http://dx.doi.org/10.1103/PhysRevB.92.060301}{Phys. Rev. B \textbf{92}, 060301(R) (2015).}

\bibitem{kurizki}  Goren Gordon, Gershon Kurizki, and Daniel A. Lidar,
\href{http://journals.aps.org/prl/abstract/10.1103/PhysRevLett.101.010403}{Phys. Rev. Lett. {\bf 101,} 010403 (2008).}

\bibitem{Chipaux} M. Chipaux, L. Toraille, C. Larat, L. Morvan, S. Pezzagna, J. Meijer, and T. Debuisschert, \href{ http://dx.doi.org/10.1063/1.4936758}{Appl. Phys. Lett. \textbf{107}, 233502 (2015).}

\bibitem{loncar2}  Linbo Shao , Ruishan Liu , Mian Zhang , Anna V. Shneidman , Xavier Audier ,
Matthew Markham , Harpreet Dhillon , Daniel J. Twitchen , Yun-Feng Xiao ,
and Marko Loncar.
\href{http://onlinelibrary.wiley.com/doi/10.1002/adom.201600039/abstract}{Advanced optical materials, {\bf 4,} 1075  (2016)}

\bibitem{loncar1}
Linbo Shao, Mian Zhang, Matthew Markham, Andrew M. Edmonds, and Marko Lon?ar,\href{http://journals.aps.org/prapplied/abstract/10.1103/PhysRevApplied.6.064008}{Physical review applied {\bf 6}, 064008 (2016)}

\bibitem{Kolkowitz} S. Kolkowitz et. al., \href{http://science.sciencemag.org/content/347/6226/1129}{Science {\bf 347} 1129 (2015)}

\bibitem{Hall2016} LT Hall et. al.,  \href{http://www.nature.com/articles/ncomms10211}{Nature Communications   {\bf  7,} 10211, (2016).}

\bibitem{Sushkov} A. O. Sushkov et. al., \href{http://pubs.acs.org/doi/abs/10.1021/nl502988n}{Nano Lett. \textbf{14}, 6443, (2014).}

\bibitem{Dominik}
Dominik Schmid-Lorch et. al., \href{http://pubs.acs.org/doi/abs/10.1021/acs.nanolett.5b00679}{Nano Lett., {\bf 15}, 4942 (2015)}

\bibitem{Kimmich} Kimmich, R., and E. Anoardo, \href{http://www.sciencedirect.com/science/article/pii/S0079656504000196}{Prog. Nucl. Magn.
Reson. Spectrosc. {\bf 44,} 257. (2004)}

\bibitem{Fanchini} F. F. Fanchini, J. E. M. Hornos, and R. d. J. Napolitano, \href{http://dx.doi.org/10.1103/PhysRevA.75.022329}{
Phys. Rev. A \textbf{75}, 022329 (2007).}

\bibitem{Bermudez1} A. Bermudez, F. Jelezko, M. B. Plenio, and A. Retzker, \href{http://dx.doi.org/10.1103/PhysRevLett.107.150503}{Phys. Rev. Lett. \textbf{107}, 150503 (2011).}

\bibitem{Bermudez2} A. Bermudez, P. O. Schmidt, M. B. Plenio, and A. Retzker, \href{http://dx.doi.org/10.1103/PhysRevA.85.040302}{Phys. Rev. A \textbf{85}, 040302(R) (2012).}


\bibitem{Cai} J.-M., Cai, F. Jelezko, N. Katz, A. Retzker, and M.B Plenio, \href{http://dx.doi.org/10.1088/1367-2630/14/9/093030}{New J. Phys. \textbf{14}, 093030 (2012).}

\bibitem{Xu} X. Xu, Z. Wang, C. Duan, P. Huang, P. Wang, Y. Wang, N. Xu, X. Kong, F. Shi, X. Rong, and J. Du,
\href{http://dx.doi.org/10.1103/PhysRevLett.109.070502}{Phys. Rev. Lett. \textbf{109}, 070502 (2012).}

\bibitem{Golter} D.A. Golter, T.K. Baldwin, H. Wang, \href{http://dx.doi.org/10.1103/PhysRevLett.113.237601}{Phys. Rev. Lett. \textbf{113}, 237601 (2014).}

\bibitem{rabl2009} P. Rabl, P. Cappellaro, M. V. Gurudev Dutt, L. Jiang, J. R.
Maze, and M. D. Lukin, \href{http://dx.doi.org/10.1103/PhysRevB.79.041302}{Phys. Rev. B {\bf 79}, 041302 (2009).}

\bibitem{Jens2010} Jens Clausen, Guy Bensky, and Gershon Kurizki, \href{http://dx.doi.org/10.1103/PhysRevLett.104.040401}{ Phys. Rev. Lett.  {\bf 104,} 040401 (2010).}


\bibitem{CaiCon} J.-M. Cai, B. Naydenov, R. Pfeiffer, L. P. McGuinness, K. D. Jahnke, F. Jelezko, M. B. Plenio and A. Retzker, \href{http://iopscience.iop.org/1367-2630/14/11/113023/article}{New J. Phys. \textbf{14}, 113023 (2012).}

\bibitem{Itsik1} I. Cohen, S. Weidt, W. K. Hensinger, and A. Retzker, \href{http://stacks.iop.org/1367-2630/17/i=4/a=043008}{New J. Phys. \textbf{17}, 043008 (2015).}

\bibitem{pathrick}  Jean Teissier, Arne Barfuss, Patrick Maletinsky, \href{https://arxiv.org/abs/1611.01515}{arXiv:1611.01515}


\bibitem{Christof} N. Timoney, I. Baumgart,	M. Johanning,	A. F. Var\'{o}n,	M. B. Plenio, 	A. Retzker	and  Ch. Wunderlich, \href{http://www.nature.com/nature/journal/v476/n7359/full/nature10319.html}{Nature \textbf{476}, 185 (2011).}

\bibitem{ADR} N. Aharon, M. Drewsen, and A. Retzker, \href{http://journals.aps.org/prl/abstract/10.1103/PhysRevLett.111.230507}{Phys. Rev. Lett. \textbf{111}, 230507 (2013).}

\bibitem{Baumgart} I. Baumgart, J.-M. Cai, A. Retzker, M. B. Plenio, and Ch. Wunderlich, \href{http://dx.doi.org/10.1103/PhysRevLett.116.240801}{Phys. Rev. Lett. \textbf{116}, 240801 (2016).}

\bibitem{Lambda} N. Aharon, I. Cohen, F. Jelezko, and A. Retzker, \href{http://dx.doi.org/10.1088/1367-2630/aa4fd3}{New J. Phys. \textbf{18}, 123012 (2016).}

\bibitem{Stark} A. Stark et. al.  arXiv preprint,
arXiv:1706.04779, 2016.


\bibitem{Baylis}
C. Baylis, M. Fellows, L. Cohen, and R. J. Marks II, \href{https://doi.org/10.1109/MMM.2014.2321253}{IEEE Microwave Mag. 15(5), 94 (2014)}.
%
%
%

\end{references}
\end{document}